\documentclass[aps,showpacs,nofootinbib]{revtex4}
\usepackage[utf8]{inputenc}
\usepackage{amsmath}
\usepackage{amsfonts}
\usepackage{mathtools}
\usepackage{comment}
\usepackage{color}
\usepackage[pdftex]{graphicx}
\usepackage{textcomp}
\usepackage{amssymb}
\usepackage{url}

\DeclarePairedDelimiter{\ceil}{\lceil}{\rceil}
\DeclareMathOperator*{\argmin}{argmin}
\DeclareMathOperator*{\argmax}{argmax}

\begin{document}

\title{Optimal information diffusion in stochastic block models}
\author{Gianbiagio Curato, Fabrizio Lillo}
\affiliation{Scuola Normale Superiore, Piazza dei Cavalieri 7, 56126 Pisa, Italy}

\date{\today}

\begin{abstract}
We use the linear threshold model to study the diffusion of information on a network generated by the stochastic block model.  
We focus our analysis on a two community structure where the initial set of informed nodes lies only in one of the two communities and we look for optimal network structures, i.e. those maximizing the asymptotic extent of the diffusion. We find that, constraining the mean degree and the fraction of initially informed nodes, the optimal structure can be assortative (modular), core-periphery, or even disassortative. We then look for minimal cost structures, i.e. those such that a minimal fraction of initially informed nodes is needed to trigger a global cascade. We find that the optimal networks are assortative but with a structure very close to a core-periphery graph, i.e. a very dense community linked to a much more sparsely connected periphery.
\end{abstract}


\maketitle

\section{Introduction}
\label{sec:Introduction}

Modeling the propagation of information (innovations, virus, fads, etc) on a network is receiving a lot of interest because of the many applications and of the interesting problems that can be posed \cite{Rogers,Adamic,Ahn}. In the formalization of this problem, the state of a node can be in two states, either informed or not. The information diffusion properties depend on (i) the microscopic model linking the state of a node with the state of its neighbors and (ii) the structure of the network. 

Concerning the first aspect, a general and successful approach \cite{Watts} has been to assume that the state of a node changes and becomes informed when more than a certain fraction of the neighbors are also informed. This means that the dynamics of the process is clearly non-linear due to the presence of the threshold \cite{Karampourniotis,Pastor-Satorras}. Since the model depends linearly on the threshold parameter, this model is termed the Linear Threshold Model (LTM). LTM was originally proposed to study collective social behavior, incorporating the idea of \textquotedblleft social reinforcement\textquotedblright. 
In economic terms, this class of problems is known as \textit{binary decisions with externalities} \cite{Schelling}.
The \textquotedblleft externality\textquotedblright represents the fact that one agent cares about the opinion of the others agents in order to take decisions. Both the detailed mechanisms
involved in binary decision problems, and also the origins of the externalities can vary widely across specific problems. LTM finds a lot of applications in economics \cite{Bikhchandani}, 
sociology \cite{Granovetter} and neural networks \cite{Gonzalez-Castillo,Misic}. 

Concerning the structure of the network, the dynamics of LTM has been extensively studied on random graphs of Erd\H{o}s-R\'{e}nyi type.
The model exhibits various critical behaviors. There is a critical threshold parameter at which a single active node can trigger a macroscopic cascade \cite{Watts}. Moreover, if the
threshold parameter is held fixed, there exists another sharp transition from an inactive state where no diffusion occurs to an active state with global diffusion, and the transition is triggered at a critical fraction 
of the initially active nodes \cite{Singh}. Also the presence of hubs and the role of degree distribution in these random graph models have been studied extensively \cite{PastorVes}



Real networks often display a structure that is not compatible with random graphs. The presence of communities has been documented in many real (social) networks and this has of course consequences on the propagation of information \cite{Onnela}. Generically, community structure within a network can facilitate information spread to a larger extent than a homogeneous random network \cite{Centola,Nematzadeh,Singh}. The impact of a community structure on the information diffusion according to the LTM has been theoretically studied very recently for the first time by Nematzadeh et al. \cite{Nematzadeh}. They considered a network described by two homogeneous modules, i.e. the internal connectivity of the two communities is the same, and the set of initially informed nodes is located in one of the two communities. The main result  is that the global diffusion can be reached when the network is not homogeneous, i.e. the presence of modules facilitates the diffusion of information from one community to the other.  Nematzadeh et al. \cite{Nematzadeh} introduced  the concept of {\it optimal modularity}: the homogeneous random graph is not optimal in order to have global informative cascades, but there exists a range of values of the modularity such that the extent of the cascade is maximal. 


However the (symmetric) modular structure is only one of the possible community structures that a network can have. A more flexible way of modeling a group structure in a network is by using the Stochastic Block Model (SBM) \cite{Holland,Karrer,Wang}. As detailed below, in an SBM the probability of a link depends only on the groups the two considered nodes belong to. One of the advantages of the SBM is that it describes not only the symmetric modular case, but also  core-periphery \cite{Csermely,Rombach,Tantari,Barucca}, assortative, and disassortative networks. 

In this paper we study the LTM dynamics on a SBM. Thus one of the objectives of the present work is to extend the analysis performed by Nematzadeh et al. to networks described by structures  richer  than the symmetric modular one.  Specifically we consider
networks described by the SBM and for simplicity we consider a network described by two equally sized communities. We introduce the concept  of \textit{optimal structure}, i.e. the parameters of the SBM which allows a global cascade.

More specifically, in the following we consider two related optimization problems. Given a value of the threshold parameter, let $\boldsymbol{p}$ be the $2\times 2$ affinity matrix of the SBM and let us assume that a fraction $\rho_1(t=0)$ of nodes in the first group is initially informed, while $\rho_2(t=0)=0$. Moreover let $\rho(t)\equiv [\rho_1(t)+\rho_2(t)]/2$ be the fraction of informed nodes at time $t$ in the network. The first optimization problem we study is
\begin{eqnarray}\label{opt1}
\boldsymbol{p}^*= \underset{\boldsymbol{p}}{\argmax} ~\rho(\infty)~~~~~~~~\textrm{s.t.}~~~~~~~\rho_1(0)=k,~~~~~~z=z_0
\end{eqnarray}
i.e. finding the SBM model with fixed mean degree $z_0$ and initial fraction $k$ of informed nodes which {\it maximizes} the fraction of informed nodes in the network at the end of the cascade. We find that the optimal structures include the modular one (in agreement with \cite{Nematzadeh}) but also include other assortative, core-periphery, and even disassortative structures. 

The second optimization problem considers the \textquotedblleft cost\textquotedblright of informing the initial set and considers the  problem
\begin{eqnarray}\label{opt2}
\boldsymbol{p}^*= \underset{\boldsymbol{p}}{\argmin}~ \rho_1(0)~~~~~~~~\textrm{s.t.}~~~~~~~\rho(\infty)=1,~~~~~~z=z_0
\end{eqnarray}
i.e. finding the SBM model with fixed mean degree $z_0$ for which a global cascade is asymptotically obtained and for which the fraction of initially informed nodes is {\it minimal}. In this case we find that the optimal structure is in the assortative regime but very close to core periphery structures, i.e. where the probability of links between the two communities is the same or slightly smaller than the one inside the initially uninformed community.




Our analysis shows the presence of tipping points similar to those observed by Singh et al. \cite{Singh} for homogeneous random networks. 
The presence of tipping points of the global density of informed nodes has also been observed by Nematzadeh et al. \cite{Nematzadeh} for modular networks.
However our analysis differs from the last one because we study separately the density of informed nodes belonging to different communities, in order to show how the
information diffuses from an informed community to an uniformed one.  
We find that the presence of tipping points is almost ubiquitous, independently from the various network structures. However the value of the critical initial density,  defining the tipping point, is strongly dependent from the network structure. 

The paper is organized as follows. In section \ref{sec:Network dynamics}
we summarize the threshold model, the stochastic block modeling approach, and how to analyze theoretically the density of informed nodes. In section \ref{sec:Cascade phase diagram} we study the onset of global cascades as a function
of the network structure by means of numerical simulations and theoretical analysis. In section \ref{sec:Critical points} we analyze the presence of tipping points as a function of the network structure and in section \ref{sec:opt} we identify the optimal network structures.
Finally in section \ref{sec:Conclusions} we draw some conclusions and perspectives of the present work.

\section{The model}
\label{sec:Network dynamics} 

Let us recap the linear threshold model \cite{Watts}. Consider a set of $N$ nodes (agents) connected by undirected edges of a network. The state of an agent $i$ at time
$t$ is described by a binary variable $s_{i}\left(t\right) \in \left\lbrace 0,1 \right\rbrace$, where $1$ represents the active state and $0$ the inactive one. At time
$t=0$ a fraction $\rho\left(0\right)$ of randomly selected agents, or seeds, is initialized in the active state. At each time step, every agent's state is updated synchronously \cite{Nematzadeh},
according to the following rule
\begin{equation}
 \label{eq:1.1}
 s_{i}\left(t+1\right)=\begin{cases}
                        1\,\,\,\textrm{if}\,\,\, \sum_{j=1}^{N}A_{ij}s_{j}\left(t\right) > \left(\theta \sum_{j=1}^{N}A_{ij}=\theta k_{i}\right) \vee s\left(t\right)=1;& \\
                        0\,\,\,\textrm{if}\,\,\, \sum_{j=1}^{N}A_{ij}s_{j}\left(t\right) \leq \left(\theta \sum_{j=1}^{N}A_{ij}=\theta k_{i}\right);&
                       \end{cases}
\end{equation}
where $A_{ij}$ is the adjacency matrix of the network, $0<\theta<1$ is the threshold parameter, and $k_{i}$ is the degree of the node $i$.
The asymptotic dynamics of this process is described by a fixed point \cite{Nematzadeh}. The linearity of the model derives from the functional dependence on the
threshold parameter, but the dynamics of the process is clearly non-linear. 

In this work we study the dynamics of LTM on a network generated by a SBM. 
The SBM was first studied in mathematical sociology by Holland et al. \cite{Holland} and by Wang and Wong \cite{Wang}. 
In the simplest SBM (several more complicated variants have been proposed, see \cite{Karrer}), each of $N$ vertices is assigned to one of $n$ blocks (or groups)
and undirected edges are placed independently between vertex pairs, with probabilities that are a function only of the group membership of the vertices. If we denote by $g_{i}$ the group to
which vertex $i$ belongs, then the $n \times n$ affinity matrix $\boldsymbol{p}$ is
such that the matrix element $p_{g_{i}g_{j}}$ is the probability of an edge between vertices $i$ and $j$. The vertex assignment is a mapping between 
integers $g_{i}:\left\lbrace 1\cdots N\right\rbrace \rightarrow \left\lbrace 1\cdots n\right\rbrace$. In the following we use the integers $i,j$ for a node pair, $r,s$ for a block pair.
$N_{r}$ denotes the size of block $r$, where $\sum_{r=1}^{n}N_{r}=N$.


When $p_{rs}=p$ for all pairs $r,s$, the SBM reduces to the Erd\H{o}s-R\'{e}nyi random graph model $G(N, p)$ \cite{Newman}. If the elements of $\boldsymbol{p}$ are not all the
same, then the SBM generates Erd\H{o}s-R\'{e}nyi random graphs within each community $r$, with an internal link density given by $p_{rr}$, and random bipartite 
graphs between pairs of communities $r$ and $s$ with link density given by $p_{rs}$. 

In this paper we study the propagation of information according to the LTM on a network generated by the SBM described by two communities, i.e $n=2$. 
The affinity matrix $\boldsymbol{p}$ is thus described by three independent probabilities, since $p_{12}=p_{21}$. 
The SBM can describe three different network structures: assortative network, disassortative network and core-periphery network. In terms of the affinity matrix, the different
cases are:
\begin{equation}
 \label{eq:1.2}
 SBM\left(\boldsymbol{p}\right)=\begin{cases}
                        \textrm{assortative if}\,\,\, \left( p_{11}>p_{12} \right)\wedge \left( p_{22}>p_{12}\right);& \\
                        \textrm{disassortative if}\,\,\, \left( p_{11}<p_{12}\right)\wedge \left( p_{22}<p_{12}\right);& \\
                        \textrm{core-periphery if}\,\,\, \left( p_{11}>p_{12}>p_{22}\right)\vee \left( p_{11}<p_{12}<p_{22} \right). &
                       \end{cases}
\end{equation}
The assortative case corresponds to a modular structure where the two blocks are internally strongly connected and in the symmetric case ($p_{11}=p_{22}$) it is the model considered in \cite{Nematzadeh}. The dissasortative structure is approximately bipartite and the majority of connections are between members of different blocks. Finally the in the core-periphery structure one densely connected block is connected with the other group, which internally is very weakly connected  

We choose the two communities of the same size, i.e. $N_{1}=N_{2}=N/2$. The propagation of the information starts at time $t=0$.
We assume that the initial set of informed nodes is located in the community $1$ with
density $\rho_{1}\left(0\right)\neq 0$, while $\rho_{2}\left(0\right)=0$. The total density of informed nodes of the network is defined by 
$\rho\left(t\right)=\left(\rho_{1}\left(t\right)+\rho_{2}\left(t\right)\right)/2$. 

We are interested to study the asymptotic state of the network state, i.e. $\rho\left(\infty\right),\,\rho_{1}\left(\infty\right),\,\rho_{2}\left(\infty\right)$, 
for a fixed value of the mean network degree $z$. In the SBM this constraint is 
\begin{equation}
 \label{eq:1.3}
 z=\left(\frac{N-2}{4}\right)p_{11}+\frac{N}{2}p_{12}+\left(\frac{N-2}{4}\right)p_{22}.
\end{equation}

\subsection{Theoretical analysis}
\label{sec:Theoretical analysis} 

In order to compute the asymptotic densities, we use two approximations: the mean field and the tree-like approximation. 

The mean field approach (MF) \cite{Nematzadeh} is a first approximation to study theoretically the asymptotic state of the density of informed nodes on the network. 
It makes use only of the information on the degree distribution, assuming that the network is otherwise random.
In the case under exam the nodes of the network are split into two groups with probabilities $\left\lbrace \alpha_{1},\alpha_{2} \right\rbrace$.
In the model considered here the two communities have the same size $N/2$, thus $\alpha_{1}=\alpha_{2}=1/2$.
Following Daudin et al. \cite{Daudin}, the conditional distribution of the degree of a node, given its group label, is binomial (approximately Poisson)
\begin{equation}
 \label{eq:1.4}
 P_{r}\left(k\right)=\mathcal{B} \left(N-1,\pi_{r}\right)\approx \mathcal{P}\left(k;\lambda_{r} \right),\,\,r\in\left\lbrace 1,2 \right\rbrace,
\end{equation}
where $\pi_{r}=\sum_{s=1}^{2}\alpha_{s}p_{rs}$ and $\lambda_{r}=\left(N-1\right)\pi_{r}$.

The mean field equations for the asymptotic densities of the informed nodes are
\begin{equation}
 \label{eq:1.5.0}
 \begin{split}
   \rho_{1,2}\left(\infty\right)= & \rho_{1,2}\left(0\right)+\left(1-\rho_{1,2}\left(0\right)\right)\sum_{k=1}^{\infty}\mathcal{P}\left(k;\lambda_{1,2} \right) \\
  & \times \sum_{m=\ceil{\theta \,k}}^{k}\binom{k}{m}\bar{\rho}_{1,2}^{m}\left(1-\bar{\rho}_{1,2}\right)^{k-m}, 
 \end{split}
\end{equation}
\begin{eqnarray}
 \label{eq:1.5} 
\ && \bar{\rho}_{1}=\frac{\alpha_{1}p_{11}}{\alpha_{1}p_{11}+\alpha_{2}p_{12}}\rho_{1}\left(\infty\right)+\frac{\alpha_{2}p_{12}}{\alpha_{1}p_{11}+\alpha_{2}p_{12}}\rho_{2}\left(\infty\right), \nonumber\\
\ && \bar{\rho}_{2}=\frac{\alpha_{2}p_{22}}{\alpha_{1}p_{12}+\alpha_{2}p_{22}}\rho_{2}\left(\infty\right)+\frac{\alpha_{1}p_{12}}{\alpha_{1}p_{12}+\alpha_{2}p_{22}}\rho_{1}\left(\infty\right). 
\end{eqnarray}

The second, more sophisticated but more accurate, approximation is the tree like approximation  \cite{Nematzadeh,Karampourniotis}. Gleeson et al. \cite{Gleeson1,Gleeson2} have derived a class of recursive equations approximating the random network by a tree structure. 
The approximation is valid only if the graph is sparse enough, i.e. the network structure is locally a tree because the presence of cycles is unlikely.
The top level of the tree is a single node of degree $k$, this is connected to its $k$ neighbors at the next lower level of the tree.
Each of these nodes is in turn connected to $k_{i}-1$ neighbors at the next lower level, where $k_{i}$ is the degree of node $i$, and so on. The degree
distribution of the nodes in the tree is a modification of the original $\mathcal{P}\left(k\right)$, i.e. it is given by $(k/z)\mathcal{P}\left(k\right)$. 
To find the final density $\rho$ of active nodes, we label the levels of the tree from $n = 0$ at the bottom, with the top node at an infinitely high level $n\rightarrow\infty$. The probability $q_{1,2}$ that a random node at level $n$ of the tree is active is given by
\begin{equation}
 \label{eq:1.6.0}
 \begin{split}
   q_{1,2}\left(n+1\right)= & q_{1,2}\left(0\right)+\left(1-q_{1,2}\left(0\right)\right)\sum_{k=1}^{\infty}\frac{k}{\lambda_{1,2}}\mathcal{P}\left(k;\lambda_{1,2} \right) \\
  & \sum_{m=\ceil{\theta \,k}}^{k-1}\binom{k}{m}\bar{q}_{1,2}\left(n\right)^{m}\left(1-\bar{q}_{1,2}\left(n\right)\right)^{k-1-m},
 \end{split}
\end{equation}
\begin{eqnarray}
 \label{eq:1.6}
\ && \bar{q}_{1}\left(n\right)=\frac{p_{11}}{p_{11}+p_{12}}q_{1}\left(n\right)+\frac{p_{12}}{p_{11}+p_{12}}q_{2}\left(n\right), \nonumber\\
\ && \bar{q}_{2}\left(n\right)=\frac{p_{22}}{p_{12}+p_{22}}q_{2}\left(n\right)+\frac{p_{12}}{p_{12}+p_{22}}q_{1}\left(n\right),  
\end{eqnarray}
with $q_{1,2}\left(0\right)=\rho_{1,2}\left(0\right)$. This system of equations allows to find the probability that the top node, i.e. that corresponding to $n\rightarrow\infty$, is active.
This probability is used as the probability of having an active node as neighbor of a given node of degree $k$. This implies different equations for
$\rho_{1,2}\left(\infty\right)$ with respect to eq. \ref{eq:1.5.0}. The asymptotic densities of informed nodes are given by
\begin{equation}
 \label{eq:1.7} 
 \begin{split}
   \rho_{1,2}\left(\infty\right)= & \rho_{1,2}\left(0\right)+\left(1-\rho_{1,2}\left(0\right)\right)\sum_{k=1}^{\infty}\mathcal{P}\left(k;\lambda_{1,2} \right) \\
  & \sum_{m=\ceil{\theta \,k}}^{k}\binom{k}{m}q_{1,2}\left(\infty\right)^{m}\left(1-q_{1,2}\left(\infty\right)\right)^{k-m}.
 \end{split} 
\end{equation}
The nonlinear systems of equations \ref{eq:1.5.0} and \ref{eq:1.6.0} are solved numerically searching for a fixed point in order to find a stable solution.

\section{Cascade phase diagram}
\label{sec:Cascade phase diagram}

In this section we investigate the presence of global cascades in the SBM for fixed $\rho_1(0)$.
The SBM considered here is characterized by the three parameters of the affinity matrix. One of them is fixed by the constraint on $z$ of eq. \ref{eq:1.3}, thus the phase space can be represented on the parametric plane $\left(Np_{11},Np_{12}\right)$. 
We use $Np_{ij}$ rather than $p_{ij}$ in order to obtain values independent from the network size in the thermodynamic limit. We remind that $p_{11}$ is the probability of connection inside the initial informed community and $p_{12}$
is the probability of connection between the two communities. 

The different network structures described by eq. \ref{eq:1.2} and the constraint on $z$ imply a division of the phase space (plane) into four
regions
\begin{eqnarray}
 \label{eq:2.1} 
\ && \left( Np_{12}>Np_{11}\right)\wedge \left( Np_{12}>\frac{4zN}{3N-2}-\frac{\left(N-2\right)N}{3N-2}p_{11}\right) \Rightarrow \textrm{disassortative} \nonumber\\
\ && \left( Np_{12}<Np_{11}\right)\wedge \left( Np_{12}<\frac{4zN}{3N-2}-\frac{\left(N-2\right)N}{3N-2}p_{11}\right) \Rightarrow \textrm{assortative} \nonumber\\
\ && \left( Np_{12}<Np_{11}\right)\wedge \left( Np_{12}>\frac{4zN}{3N-2}-\frac{\left(N-2\right)N}{3N-2}p_{11}\right) \Rightarrow \textrm{core-periphery} \nonumber\\
\ && \left( Np_{12}>Np_{11}\right)\wedge \left( Np_{12}<\frac{4zN}{3N-2}-\frac{\left(N-2\right)N}{3N-2}p_{11}\right) \Rightarrow \textrm{periphery-core} \nonumber\\
\end{eqnarray}
The conditions on the right of the $\wedge$ correspond to $p_{ij}>0$, which implies $Np_{12}<2z-\left(N-2\right)p_{11}/2$. 
Finally the internal density of the first block is larger than the one in the second block when
\begin{equation}
 \label{eq:2.2} 
 p_{11}>p_{22}\,\Rightarrow\, Np_{12}>2z-\left(N-2\right) p_{11}.
\end{equation}

\begin{figure}[t]
	\centering
		\includegraphics[width=1\textwidth]{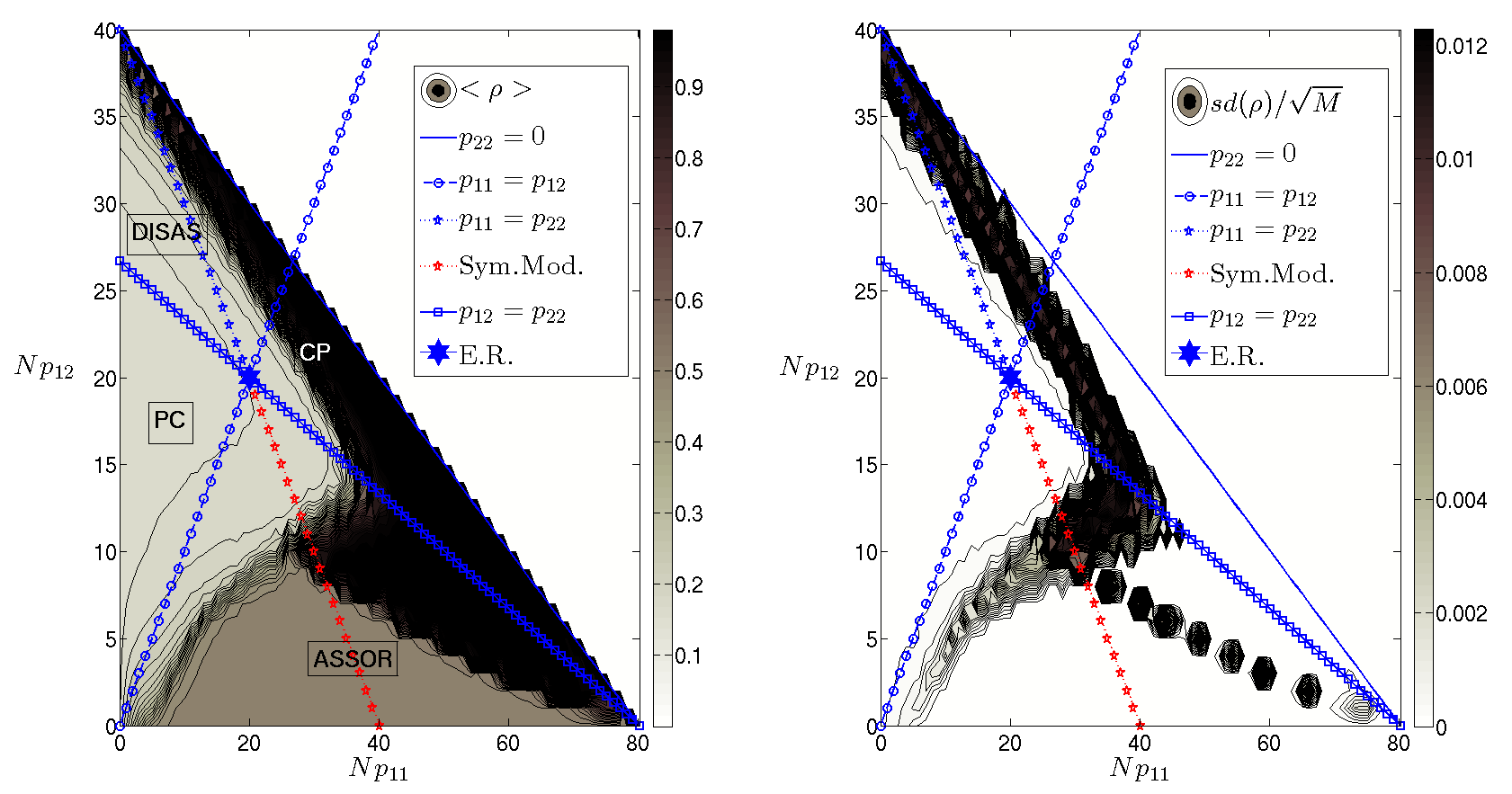} 
	\caption{(Color online). Left panel: Phase diagram for the mean value of the density $\rho\left(\infty\right)$ obtained simulating the LTM dynamics in the case: $N=1,000, n=2, z=20, \theta=0.4, \rho_{1}\left(0\right)=0.34$.
	         Right panel: Standard error of the mean. The hexagram stands for the Erd\H{o}s-R\'{e}nyi graph. The star-dotted red segment  represents the parametric region
	          studied by Nematzadeh et al. \cite{Nematzadeh}, i.e. the symmetric modular structure (Sym. Mod). 
	          We use a mesh $N\Delta p_{11}=N\Delta p_{12}=1$.} 
	\label{fig:Phase_tot_theta_04_new_2}
\end{figure}

The phase space is therefore divided into four regions (see Eq. \ref{eq:2.1}) as shown in fig. \ref{fig:Phase_tot_theta_04_new_2}.  Notice that the core-periphery regions are indicated by the acronyms PC and CP. The first one represents the case where the initial informed nodes are inside the core block, while the second one represents the case where the initial informed nodes are in the periphery.

The region where $\rho(\infty)=1$ corresponds to global cascade, therefore we search SBM structures such that, for the given mean degree $z$ and fraction $\rho_1(0)$ of initially informed nodes, lead to a global cascade. In other words we are looking for solutions of the optimization problem of Eq. \ref{opt1}. Nematzadeh et al. \cite{Nematzadeh} have investigated the same system only on the region of parameters defined by the relation $p_{11}=p_{22}$ and $p_{12}<p_{11}$, i.e. the assortative case where two modules have the same connectivity. In the phase space this is a segment indicated by star-dotted red line in Fig.  \ref{fig:Phase_tot_theta_04_new_2}. Our contribution is therefore to extend greatly the set of networks considering structures radically different from the symmetric modular one. 

We have studied the informative cascades simulating the LTM dynamics on networks of size $N=1,000$ using $M=100$ different realizations of the SBM for many points of the plane $\left(Np_{11},Np_{12}\right)$ \footnote{The results regarding the asymptotic density does not change substantially for $N=10,000$, however
there is a reduction of the mean error. 
}. The parameters are $z=20$, $\theta=0.4$ and $\rho_{1}\left(0\right)=0.34$. These parameters are the same as those used in \cite{Nematzadeh}. The first two will be kept fixed in the rest of the paper, while in the next sections we vary the latter one.
Fig. \ref{fig:Phase_tot_theta_04_new_2} shows the mean value and the standard error
of the asymptotic density $\rho(\infty)$ computed on the different realizations of the same SBM network. 

We observe that the phase space is divided in three phases: the first one (light brown) is characterized by $\rho\left(\infty\right)\approx0.17=\rho_1(0)/2$ describes the case when the cascade does not propagate neither in group 1 nor in group 2. The second one (dark brown) corresponds to $\rho\left(\infty\right)\approx0.5$ describing the case when the cascade invades all the initially informed group 1, but does not propagate into group 2 ($\rho_{B}\left(\infty\right)=0$). Finally the third phase (black), characterized by $\rho\left(\infty\right)=1$, describes the case of a global cascade where asymptotically all the nodes are informed.

Our results show that the optimal structure ($\rho\left(\infty\right)=1$) has a complex shape, i.e. it is not described by a single specific network structure. The optimal region crosses the disassortative, core-periphery (CP), and assortative regions. In the case of 
fig. \ref{fig:Phase_tot_theta_04_new_2} if the propagation of information started in the periphery (the PC region), the core would always be asymptotically uniformed\footnote{This depends on the chosen parameters and in particular on $\theta$. If this parameter is small enough (e.g. $\theta=0.2$)  the global cascade phase includes also the PC region.}.  The separation between the phases is sharp and fluctuation analysis (right panel of fig. \ref{fig:Phase_tot_theta_04_new_2}) indicates that results are very stable with the expected exception of large fluctuations very close to the boundaries between phases. Finally the non monotonic behavior of $\rho\left(\infty\right)$ as a function of the fraction of links between the communities documented in \cite{Nematzadeh} is here reproduced. In fact, by increasing this fraction, i.e. moving upward along the red line, the system goes from the phase with cascades only in group 1, to the global cascade phase, and finally back to the non propagation phase, until the Erdos-Renyi configuration is reached. It is interesting to note that, continuing along the line in the disassortative region (a case not considered in  \cite{Nematzadeh}), for $p_{12}$ large enough the system goes back to the global cascade phase. This case corresponds to an almost bipartite structure and, to the best of our knowledge, this is the first documentation of a global cascade phase of the linear threshold model in an (almost) bipartite system.

\begin{figure}[t]
	\centering
		\includegraphics[width=1\textwidth]{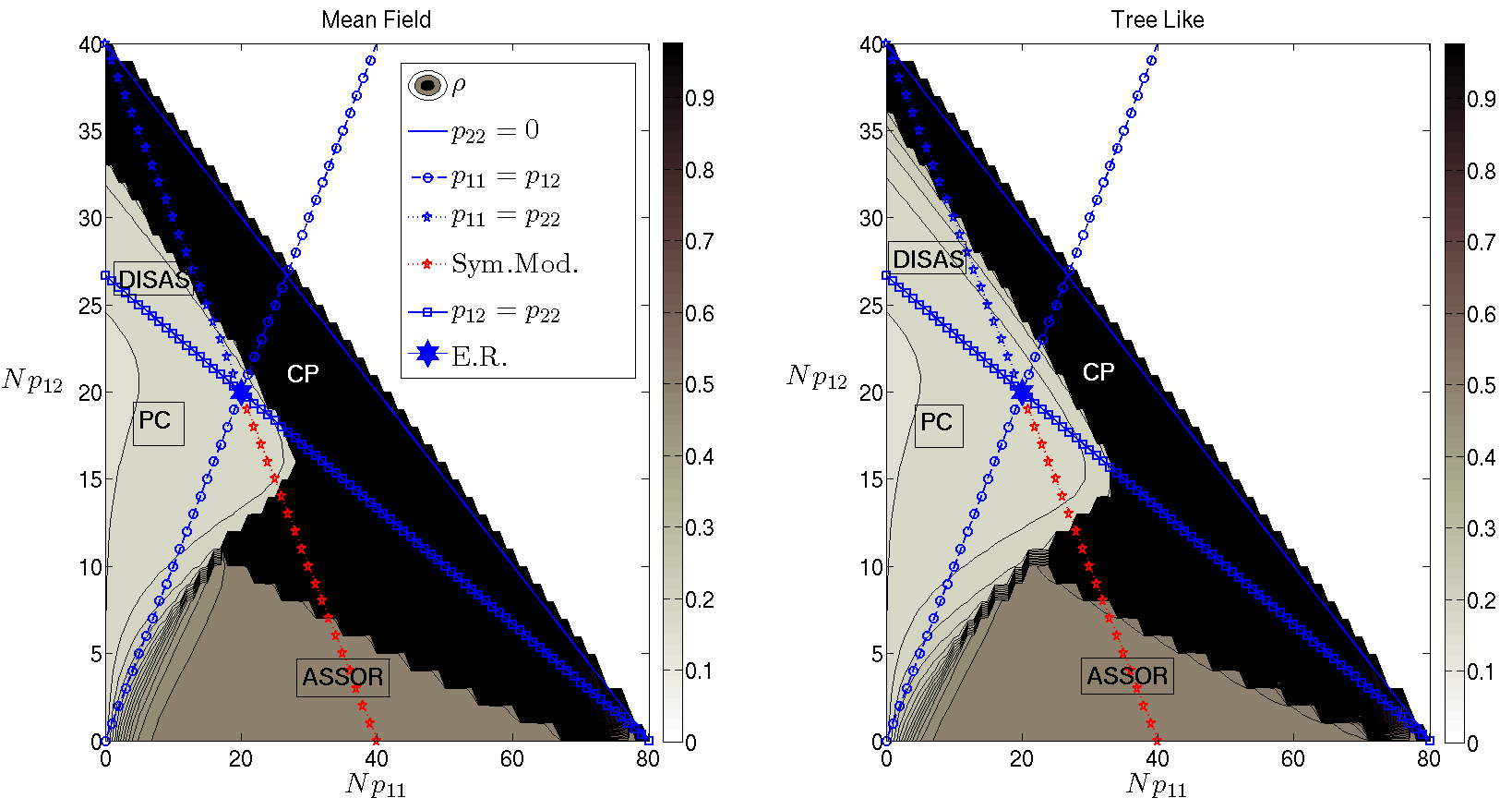} 
	\caption{(Color online). Left panel: Phase diagram describing the asymptotic density $\rho\left(\infty\right)$ computed from the mean field approximation of eq. \ref{eq:1.5.0}.
	         Right panel: Phase diagram describing $\rho\left(\infty\right)$ computed from the tree like approximation of eq. \ref{eq:1.7}. The parameters are the same as those described in the caption of Fig. \ref{fig:Phase_tot_theta_04_new_2}. }
	\label{fig:Rho_MF_TL} 
\end{figure}

The results of the simulations are confirmed by the theoretical analysis presented  in sec. \ref{sec:Theoretical analysis}, as we can observe in fig. \ref{fig:Rho_MF_TL}. 
As already observed by Nematzadeh et al. \cite{Nematzadeh}, the mean-field approximation overestimates the presence of global cascades with respect to the 
simulation results. The tree-like approximation, instead, is a very good approximation of simulation results.

\section{Tipping points}
\label{sec:Critical points}

In this section we analyze in more details the onset of informative cascades. First, we analyze the asymptotic density of informed nodes as a function
of the initial density of informed nodes $\rho_{1}\left(0\right)$ in order to find critical tipping points, i.e. critical initial densities for the onset of the global cascade. Then, we investigate the value of these initial critical densities as a function of the phase space $\left(Np_{11},Np_{12}\right)$. 

\begin{figure}[t]
	\centering
		\includegraphics[width=0.85\textwidth]{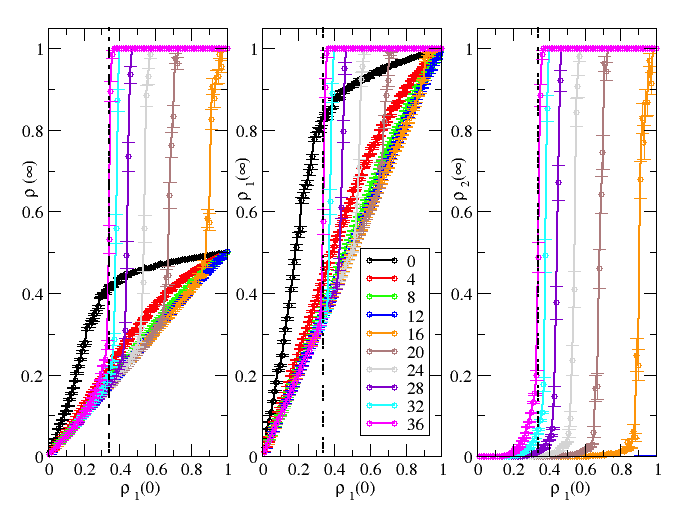} 
	\caption{(Color online). Asymptotic densities $\rho\left(\infty\right)$, $\rho_{1}\left(\infty\right)$, $\rho_{2}\left(\infty\right)$ as a function
	         of the initial density of informed nodes $\rho_{1}\left(0\right)$, for the case $N=1,000, n=2, z=20, \theta=0.4$.
	         The legend reports the values of the parameter $Np_{12}$. The internal connectivity of group 1 is constant: $Np_{11}=4$.  
	         We use a mesh $N\Delta p_{11}=N\Delta p_{12}=4$. The dark vertical dotted line indicates the value $\rho_{1}\left(0\right)=0.34$, 
	         corresponding to asymptotic densities showed in fig. \ref{fig:Phase_tot_theta_04_new_2}.}
	\label{fig:Asym_Density_sim_N_1000_2} 
\end{figure}

The presence of tipping points has been extensively studied on homogeneous random
graphs \cite{Singh}. We consider here tipping points for an heterogenous system (SBM) and therefore we can observe tipping points relative to the global network and to the single communities. 
Moreover Singh et al. \cite{Singh} performed simulations of the LTM dynamics on empirical networks composed by two communities, documenting that sometimes the cascade first sweeps
one of the communities, while the other resists before the cascade becomes global. Here instead we select the initially informed
nodes only in one community in order to observe the transmission of information from one informed community to the other uniformed one. 

We analyze the asymptotic densities by fixing the values of the internal density, i.e. $Np_{11}=4,20$, and varying the density of the connections between communities, $Np_{12}$, in order to investigate the presence of tipping points. We choose these values because in the second case we always observe tipping points, while in the first case this is not always true.  We report the results given by simulations. The tree like approximation confirms such results (not shown).  

\begin{figure}[t]
	\centering
		\includegraphics[width=0.85\textwidth]{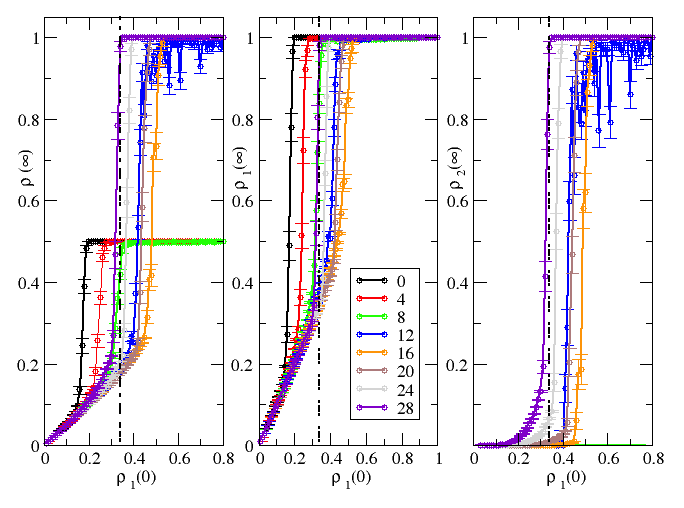} 
	\caption{(Color online). Asymptotic densities $\rho\left(\infty\right)$, $\rho_{1}\left(\infty\right)$, $\rho_{2}\left(\infty\right)$ as a function
	         of the initial density of informed nodes $\rho_{1}\left(0\right)$, for the case $N=1,000, n=2, z=20, \theta=0.4$.
	         The legend reports the values of the parameter $Np_{12}$. The internal connectivity of group 1 is constant: $Np_{11}=20$.  
	         We use a mesh $N\Delta p_{11}=N\Delta p_{12}=4$. The dark vertical dotted line indicates the value $\rho_{1}\left(0\right)=0.34$, 
	         corresponding to asymptotic densities showed in fig. \ref{fig:Phase_tot_theta_04_new_2}.}
	\label{fig:Asym_Density_sim_N_1000} 
\end{figure}

Fig. \ref{fig:Asym_Density_sim_N_1000_2} shows the asymptotic densities when $Np_{11}=4$ for ten allowed values of $Np_{12}$. The dependence of $\rho\left(\infty\right)$ as a function of $\rho_{1}\left(0\right)$ can be described either by a smooth function
or by a discontinuous one, indicating the presence of a tipping point. For values $Np_{12}\leq12$ we do not detect sharp transitions of the global
density $\rho\left(\infty\right)$. The informative cascade propagates only inside the initially informed community for each value of $\rho_{1}\left(0\right)$. 
When $Np_{12}>12$ the information can propagate inside the second community. The asymptotic density of the second community shows always a tipping point describing
a sharp transition from a state where we have substantially zero information to a state where the community is fully informed. This is the main difference
with the behavior of the first community, where the density of information gradually increases from zero and than make a jump described by a tipping point. 
The tipping points are the same for the two communities

The behavior of the densities of the global network and of the initially informed community are similar to that described by Singh et al. \cite{Singh} for homogeneous random networks, 
but the behavior of the initially uniformed community was not known to the best of our knowledge. This last behavior is similar to the resistance effect described by Singh et al. \cite{Singh}
for an empirical high school friendship network described by two communities.
It manifests itself when the initial information belongs only to one community and the information diffuses to the uniformed community in a strong nonlinear 
way in terms of $\rho_{1}\left(0\right)$.

\begin{figure}[t]
	\centering
		\includegraphics[width=1\textwidth]{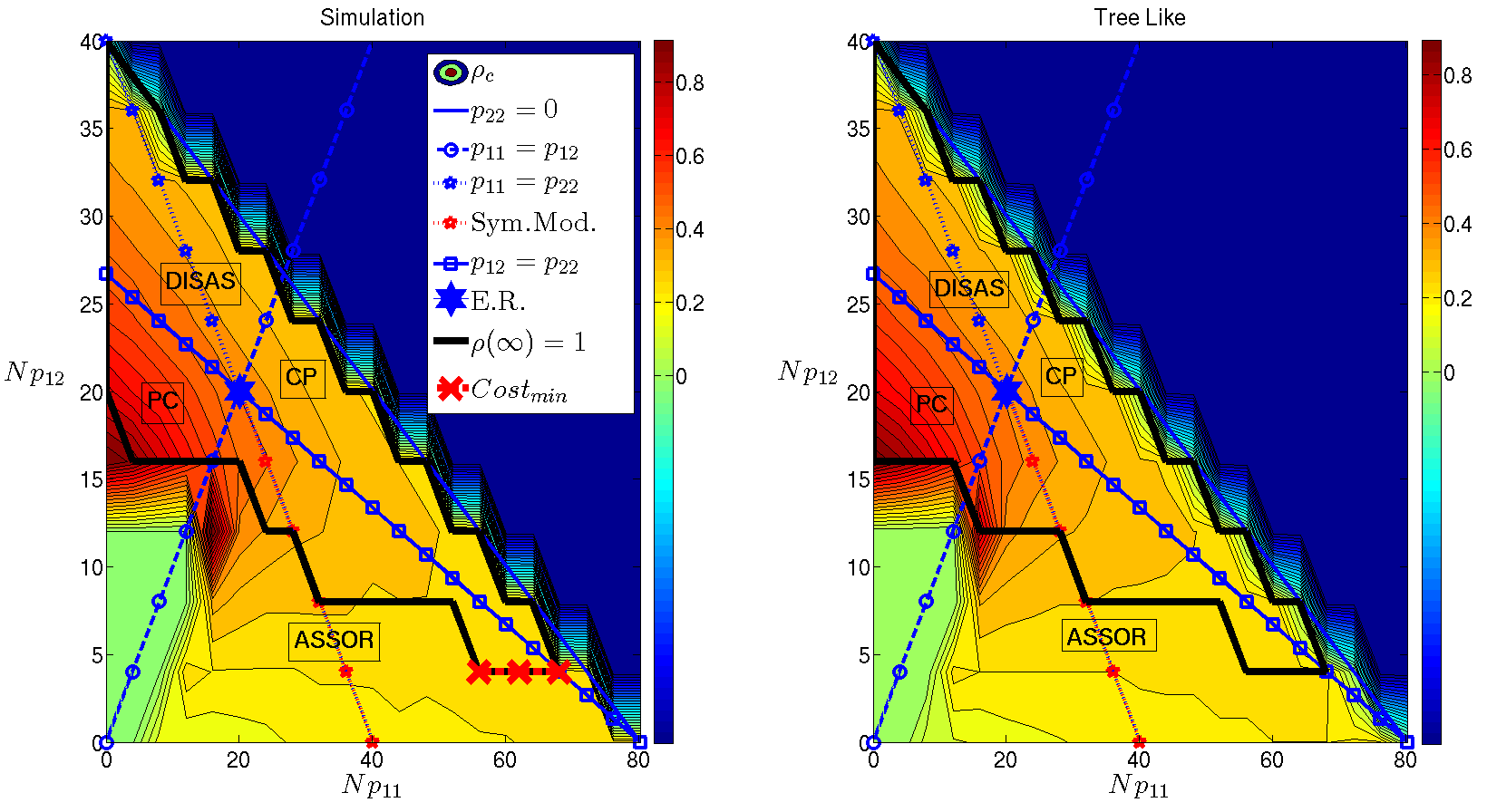} 
	\caption{(Color online). Values of the tipping points $\rho_{c}$ of the initial density of group 1 for $\rho\left(\infty\right)$ as a function of the network structure. We report results relative to the simulations in the case
	        $N=1,000, n=2, z=20, \theta=0.4$ and the values obtained solving numerically the tree-like equations. The green region
	        indicates the absence of tipping points. Dark blue values above the main diagonal indicate the not feasible region where $p_{22}<0$. The area bounded by the black thick line corresponds to SBMs where a global cascade ($\rho(\infty)=1$) is observed. Finally the red crosses ($Cost_{min}$) correspond to the optimal structures.}
	\label{fig:Tipping_Rho_RhoA_2} 
\end{figure}

\begin{figure}[t]
	\centering
		\includegraphics[width=1\textwidth]{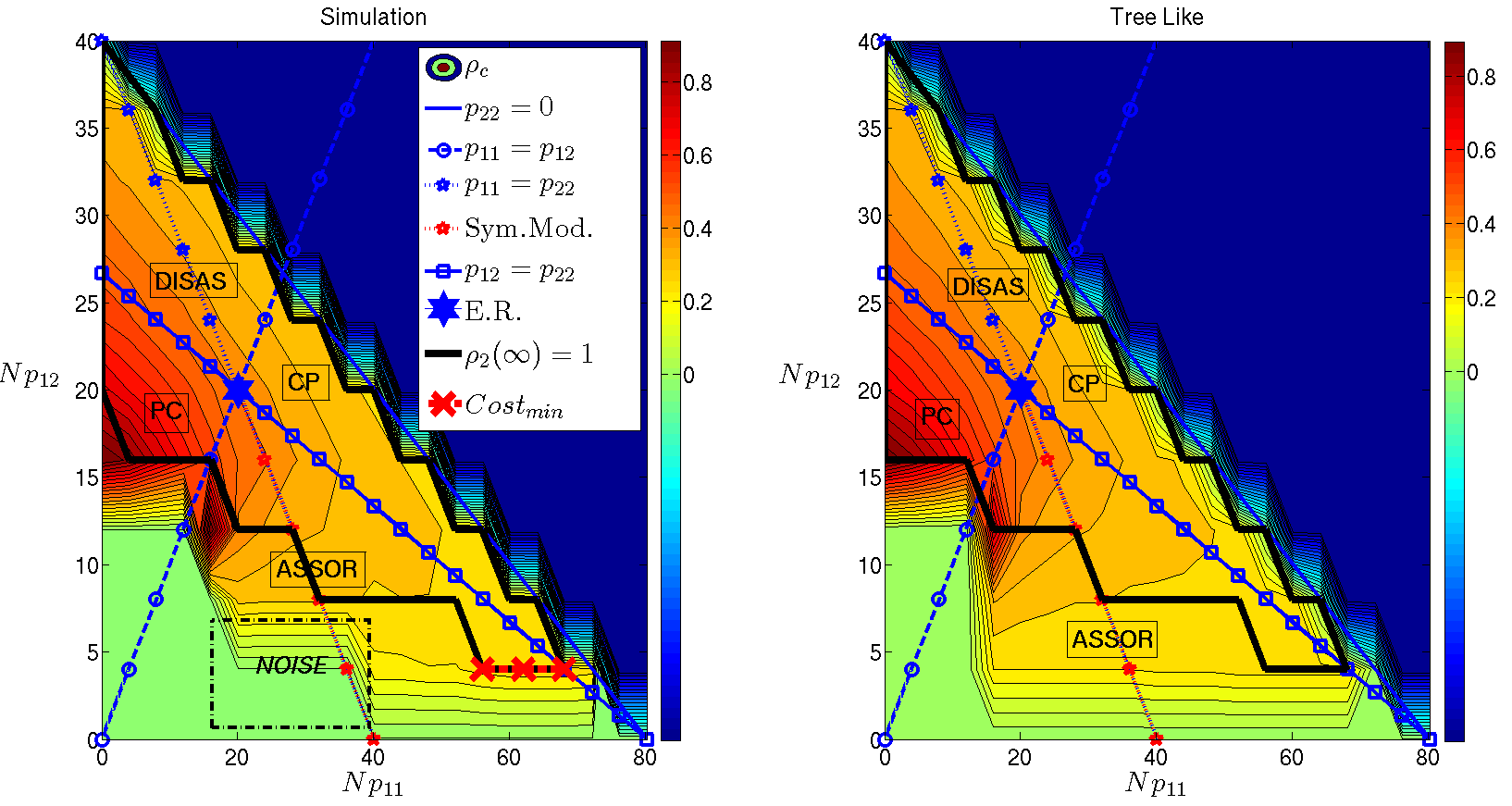} 
	\caption{(Color online). Values of the tipping points $\rho_{c}$ of the initial density of group 1 for $\rho_{2}\left(\infty\right)$ as a function of the network structure. We report results relative to the simulations in the case
	        $N=1,000, n=2, z=20, \theta=0.4$ and the values obtained solving numerically the tree-like equations. The noise region represents a zone with densities compatible
	        with zero values. The green region indicates the absence of tipping points. Dark blue values above the main diagonal indicate the not feasible region where $p_{22}<0$. The area bounded by the black thick line corresponds to SBMs where a global cascade ($\rho(\infty)=1$) is observed. Finally the red crosses ($Cost_{min}$) correspond to the optimal structures.}
	\label{fig:Tipping_new_RhoB_RhoA_2}  
\end{figure}

 Fig. \ref{fig:Asym_Density_sim_N_1000} shows the asymptotic densities when $Np_{11}=20$ for ten allowed values of $Np_{12}$. In this case 
the dependence of $\rho\left(\infty\right)$ with respect to $\rho_{1}\left(0\right)$ is always described by sharp transitions. When $Np_{12}=4,8$ our simulations of LTM give a quite noisy density  $\rho_{2}\left(\infty\right)$, i.e. the errors are compatible with a zero density\footnote{The tree like equation shows a sharp transition
to an asymptotic state of $\rho_{2}\left(\infty\right)\approx0.002$ for $Np_{12}=8$ and $\rho_{2}\left(\infty\right)\approx0.00001$ for $Np_{12}=4$. These results may indicate
that for such kind of asymptotic values the effects of finite size, e.g. $N=1,000$, on the LTM dynamics become dominant and simulations with a higher numbers of nodes are required.}.
For values of $Np_{12}>8$ we can observe again sharp transitions for the density $\rho_{2}\left(\infty\right)$. The critical points of $\rho_{1}\left(\infty\right)$ 
and $\rho_{2}\left(\infty\right)$ are the same outside of the noisy region. 

\section{Optimal structures}\label{sec:opt}

We now consider the optimization problem stated in Eq. \ref{opt2}. If informing each initial node has a cost (and the mean degree is fixed), which is the SBM network structure which requires the minimal cost in order to reach a global cascade? 

To answer this question we first investigate the position of the critical points (when present) as a function of the parameters of the SBM.
The critical points 
$\rho_{c}$ for $\rho_1(0)$ are calculated numerically by computing the derivative of $\rho\left(\infty\right)$, $\rho_{1}\left(\infty\right)$ (not shown below) and $\rho_{2}\left(\infty\right)$ 
with respect to $\rho_{1}\left(0\right)$ and finding its maximum.  Fig. \ref{fig:Tipping_Rho_RhoA_2} and fig. \ref{fig:Tipping_new_RhoB_RhoA_2} show the values of the tipping points $\rho_{c}$ for the global network and the initially uniformed
community, respectively, on the phase space $\left(Np_{11},Np_{12}\right)$. 
The critical points occur for the same values in the two communities except for the region close to $Np_{12}=0$.
The values of $\rho_{c}$ observable in figs. \ref{fig:Asym_Density_sim_N_1000_2} and \ref{fig:Asym_Density_sim_N_1000} can be inferred from figs. \ref{fig:Tipping_Rho_RhoA_2} 
and \ref{fig:Tipping_new_RhoB_RhoA_2} for $Np_{11}=4,20$. The case $Np_{11}=4$ is an example of the region where there are no tipping points near to the origin $\left(0,0\right)$. 
The presence of tipping points is instead the common behavior on the plane when $Np_{12}=20$. The behavior of the values of critical points is quite complex. 
The greatest values for the critical values belong to the PC region. 
This result is expected, because it implies that a large initial density of initiators in the periphery is required in order the information can penetrate in the core of the network.

Once the value of critical points have been obtained we superimpose in Fig.s \ref{fig:Tipping_Rho_RhoA_2} and  \ref{fig:Tipping_new_RhoB_RhoA_2} the region where the transition leads to a global cascade. This is the area bounded by the black thick line. Outside this region a transition is observed but the asymptotic state is $\rho(\infty)<1$. From an optimization point of view, inside this region one has to look for the points where the $\rho_1(0)$ is minimal.  These values correspond to the case where $Np_{11}\in [56,68]$ and $Np_{12}=4$ and in this case $\rho_{c}=0.26$\footnote{Note that we used a mesh resolution of $0.01$ in the search of $\rho_{c}$, hence these structures are equivalent within this resolution.}. These structures are optimal, since, given the threshold and the mean degree, a minimal fraction $\rho_c$ of informed nodes are needed to create a global cascade in the networks. The optimal structures are assortative but, they are far from the symmetric modular regime ($p_{11}=p_{22}$), and they have a small link probability between the two communities. Also in the limit when $p_{12}=p_{22}$, i.e.  the probability of links between the two communities is the same as the probability of links inside the initially uninformed community, the structure is optimal. 

Finally the right panels of Fig. \ref{fig:Tipping_Rho_RhoA_2} and fig. \ref{fig:Tipping_new_RhoB_RhoA_2}  show also the result for the tree like approximation. In general the agreement with numerical simulations is very good. As before, the mean-field approximation is also quite close, even if less accurate than the tree like approximation (data not shown).

\section{Conclusions}
\label{sec:Conclusions}
In this work we have used the LTM to study the information diffusion on a network described by a SBM. This extends the results of Nematzadeh et al. \cite{Nematzadeh} on symmetric modular networks and of Singh et al. \cite{Singh} relative to the LTM dynamics on random networks. We have analyzed the LTM
dynamics on a SBM network described by two communities. The initial set of informed nodes is located only in one community. This choice allows us to study how the information
mediated by social reinforcement is transmitted to an initial uniformed community. We perform such analysis by using both simulations and analytic approaches. 

We have introduced the concept of \textit{optimal network structure} as those structures that can lead to global cascades. For a fixed fraction of initially informed nodes, the optimal structure describes a shape in the two dimensional phase diagram, which includes core-periphery, assortative, and even disassortative structure.

We have studied the presence of critical tipping points for the information density of the initial uniformed community.
The effect of resistance to information transmission observed empirically by Singh et al. \cite{Singh} is confirmed by our results. The common behavior
is described by a sharp transition of the information density, in terms of the density of initiators in the informed community, from an initial uniformed
state to a global informed state. The behavior of the value of critical densities is analyzed as a function of the two dimensional parametric space described above. Both phenomena are described coherently by simulations and analytic approaches, i.e. mean field and tree-like approximations. The analysis of tipping points is useful to identify optimal structures as those where a minimal number of initially informed nodes is needed to have a global cascade. We have found that the solution of this optimization are network structures with assortative mixing but very close to a core-periphery structure. 

Such results can be useful to analyze the diffusion of information in networks described by different communities. 
The extension of this analysis to different network structures, i.e. different from SBM or with more communities, is left for future research.

\section{Acknowledgments}
This research activity was supported by Unicredit S.p.a. under the project: \textquotedblleft Dynamics and
Information Research Institute - Quantum Information (Teoria dell'Informazione), Quantum
Technologies \textquotedblright. FL acknowledges support by the European Community's H2020 Program under the scheme INFRAIA-1- 2014-2015: Research Infrastructures, grant agreement no. 654024 SoBigData: Social Mining \& Big Data Ecosystem

\end{document}